\documentclass{elsart}

\usepackage{indentfirst}
\usepackage{graphicx}
\usepackage{psfrag}

\usepackage{subfigure}

\usepackage[english]{babel}
\usepackage[latin1]{inputenc}

\usepackage[sort,sectionbib]{natbib}

\usepackage{xspace} 
\usepackage{amsmath,amssymb}

\usepackage{color}
\definecolor{oneblue}{rgb}{0,0.0,0.75}
\usepackage[colorlinks,
            urlcolor=oneblue,
            linkcolor=oneblue,
            citecolor=oneblue,
            bookmarksopen=false,
            pagebackref]{hyperref}

\journal{Physics Letters A}

\newtheorem{remark}{Remark}

\newcommand{\pd}[2]{\frac{\partial#1}{\partial#2}}
\newcommand{\abs}[1]{\left|#1\right|}
\newcommand{\set}[1]{\left\{#1\right\}}

\renewcommand{\Re}{\mathop{\mathrm{Re}}}
\renewcommand{\Im}{\mathop{\mathrm{Im}}}

\def\grad{\nabla}
\def\R{\mathbb{R}}
\def\U{\mathbb{U}}
\def\x{\vec{x}}
\def\k{\vec{k}}

\begin{document}

\begin{frontmatter}

\title{Group and phase velocities in the free-surface visco-potential flow: new kind of boundary layer induced instability}
\author[D. Dutykh]{\href{http://www.lama.univ-savoie.fr/~dutykh}{Denys Dutykh}\corauthref{cor}}
\address[D. Dutykh]{LAMA, UMR5127 CNRS, Universit\'e de Savoie, 73376 Le Bourget-du-Lac Cedex, France}
\ead{Denys.Dutykh@univ-savoie.fr}
\corauth[cor]{Corresponding author.}

\begin{abstract}
Water wave propagation can be attenuated by various physical mechanisms. One of the main sources of wave energy dissipation lies in boundary layers. The present work is entirely devoted to thorough analysis of the dispersion relation of the novel visco-potential formulation. Namely, in this study we relax all assumptions of the weak dependence of the wave frequency on time. As a result, we have to deal with complex integro-differential equations that describe transient behaviour of the phase and group velocities. Using numerical computations, we show several snapshots of these important quantities at different times as functions of the wave number. Good qualitative agreement with previous study \cite{Dutykh2008a} is obtained. Thus, we validate in some sense approximations made anteriorly. There is an unexpected conclusion of this study. According to our computations, the bottom boundary layer creates disintegrating modes in the group velocity. In the same time, the imaginary part of the phase velocity remains negative for all times. This result can be interpreted as a new kind of instability which is induced by the bottom boundary layer effect.

\end{abstract}


\end{frontmatter}

\maketitle

\section{Introduction}

The classical potential free-surface flow theory is known to be a good and relatively inexpensive model of water waves (especially in comparison with free-surface Navier-Stokes equations formulation \citep{Harlow1965, Scardovelli1999, Wu2007}). However, there are some physical situations where viscous effects cannot be neglected. The necessity of including some dissipation into various water waves models was pointed out explicitly in a number of experimental studies \citep{Zabusky1971, Wu1981, Bona1981}. For example, in the ``Résumé'' section of \citep{Bona1981} one finds:
\begin{quote}
  \dots it was found that the inclusion of a dissipative term was much more important than the inclusion of the nonlinear term, although the inclusion of the nonlinear term was undoubtedly beneficial in describing the observations\dots
\end{quote}
Obviously this conclusion is related to dissipation description only and does not have the general character. One can find many other evidences in the literature which point out the importance of viscous effects.

Historically, the researchers tried first to include the dissipative effects into various long wave models such as Burgers, Korteweg-de Vries and Boussinesq equations. There is a vast literature on this subject \citep{Keulegan1948, Ott1970, KM, Miles1976, Matsuuchi1976, Khabakhpashev1987, Sugimoto1991, Khabakhpashev1997, Dutykh2007, Dutykh2007a}.

In order to include some dissipation into the framework of free-surface potential flows, we developed the so-called visco-potential formulation \citep{Liu2004, Dias2007, DutykhDias2007, Dutykh2007a, Dutykh2008a}. The kinematic viscosity (or eddy viscosity more precisely) appears through local dissipative terms in kinematic and dynamic free-surface boundary conditions. The main peculiarity consists in modifying the bottom kinematic condition due to the presence of the boundary layer which is assumed to be laminar. Mathematically, this correction procedure leads to a nonlocal in time term. Note, that from fractional calculus point of view this nonlocal term is also a half-order integral. The physical relevance of the visco-potential formulation was shown in \citep{Liu2006}. They compared model predictions with experiments on the damping and shoaling of solitary waves. It was shown that the viscous damping due to the bottom boundary layer is well represented by this theory.

To complete our literature review, recall that there is also an alternative approach to potential flows of viscous fluids developed by Daniel Joseph and his collaborators \citep{Joseph1994, Funada2002, Joseph2004, Joseph2006}.

The main goal of the present article is twofold. On one hand, we refine our previous dispersion relation analysis. In particular, we do not neglect the evolution terms $it\pd{\omega}{t}$, where $\omega(t; \k)$ is the wave frequency (\ref{eq:plane}). Consequently, we have to deal with complex integro-differential equations. On the other hand, we show that bottom boundary layer can induce a disintegrating instability of the wave packets. This result is new to author's knowledge.

The present article is organized as follows. In Section \ref{sec:math} we describe the governing equations and perform classical dispersion relation analysis. Then, we derive an equation for the group velocity in Section \ref{sec:group}. Section \ref{sec:num} contains several numerical results and their discussion. Finally, this article is ended by outlining main conclusions of the study in Section \ref{sec:concl}.

\section{Mathematical formulation and dispersion relation analysis}\label{sec:math}

Consider the 3D fluid domain bounded above by the free-surface $z = \eta(\x,t)$, $\x = (x,y)$ and below by the rigid boundary $z = -h(\x)$. A Cartesian coordinate system with the $z$-axis pointing vertically upwards and the $xOy$-plane coinciding with the still-water level. The flow is assumed incompressible and the fluid viscous with the kinematic viscosity $\nu$. The governing equations of the free-surface visco-potential flow have the following form \citep{DutykhDias2007, Dutykh2007a, Dutykh2008a}:
\begin{equation}\label{eq:gov1}
  \Delta\phi = 0, \qquad (\x,z) \in \Omega = \R^2\times[-h,\eta],
\end{equation}
\begin{equation}
  \eta_{t} + \grad\eta\cdot\grad\phi = \phi_{z} + 2\nu\Delta\eta, \qquad z=\eta,
\end{equation}
\begin{equation}
  \phi_{t} + \frac12\abs{\grad\phi}^2 + g\eta + 2\nu\phi_{zz} = 0, \qquad z=\eta,
\end{equation}
\begin{equation}\label{eq:gov2}
  \phi_{z} = -\sqrt{\frac{\nu}{\pi}} \int\limits_0^t\frac{\phi_{zz}(\x, z=-h, \tau)}{\sqrt{t-\tau}}\; d\tau, \qquad z = -h.
\end{equation}
We can derive corresponding long wave models from the visco-potential formulation (\ref{eq:gov1}) -- (\ref{eq:gov2}). The derivation can be found in \citep{Liu2004, DutykhDias2007, Dutykh2007a}.

In order to perform the dispersion relation analysis, we have to linearize equations (\ref{eq:gov1}) -- (\ref{eq:gov2}) over the flat bottom $z = -h$. This procedure is classical \citep{Stoker1957, Whitham1999} and we do not detail it here. Then, we look for the following periodic plane wave solutions:
\begin{equation}\label{eq:plane}
  \phi(\x,z,t) = \varphi(z) e^{i(\k\cdot\x-\omega(t;\k) t)}, \qquad
  \eta(\x,t) = \eta_0 e^{i(\k\cdot\x-\omega(t;\k) t)},
\end{equation}
where $\k$ is the wavenumber and $\omega(t;\k)$ is the wave frequency.
\begin{remark}
  It is important to assume from the beginning of the derivation that the wave frequency $\omega$ explicitly depends on the time $t$. Consequently, we get some additional important terms of the form $it\pd{\omega (t; \k)}{t}$ (see equation (\ref{eq:omega})) which were neglected in our previous study \citep{Dutykh2008a}.
\end{remark}

We plug the special form of solutions (\ref{eq:plane}) into the linearized version of the governing equations (\ref{eq:gov1}) -- (\ref{eq:gov2}). After performing some simple computations (details can be found in \citep{Dutykh2007a, Dutykh2008a}), we come to the following necessary condition of periodic solution (\ref{eq:plane}) existence:
\begin{equation}\label{eq:omega}
  \Omega^2 + gk\tanh(kh) - kF(\omega,t)(\Omega^2\tanh(kh) + gk) = 0,
\end{equation}
where we introduced several notations $k := |\k|$, $\Omega := it\partial_t\omega + i\omega - 2\nu k^2$ and $F(\omega)$ is inherited from the nonlocal term:
\begin{equation*}
  F(\omega,t) := \sqrt{\frac{\nu}{\pi}}\int\limits_0^t
  \frac{e^{i\omega(\tau)(t-\tau)}}{\sqrt{t-\tau}}\;d\tau.
\end{equation*}
Once the wave frequency $\omega$ is computed from equation (\ref{eq:omega}), the phase speed can be immediately deduced by its definition:
\begin{equation*}
  c_p (t;k) := \frac{\omega(t;k)}{k}.
\end{equation*}

\begin{remark}
  The classical dispersion relation for water waves $\omega^2 = gk\tanh(kh)$ can be immediately recovered from (\ref{eq:omega}) if we replace $\Omega$ by $i\omega$ and $F(\omega)$ by $0$.
\end{remark}

\subsection{Group velocity}\label{sec:group}

The group velocity is defined as follows: 
\begin{equation*}
  c_g (t; k) := \pd{\omega(t; k)}{k}.
\end{equation*} 
There are several physical interpretations of this quantity. One can see it as the wave energy propagation speed. Another interpretation consists in viewing it as the wavetrain amplitude variation speed. Anyhow, the group speed $c_g$ plays an important r\^ole in the description of the wavetrains and wavepackets. The evolution equation of this quantity can be obtained by simple differentiation of (\ref{eq:omega}) with respect to the wavenumber modulus $k$:
\begin{multline}\label{eq:group}
  2\Omega\Gamma + g\tanh(kh) + gkh(1-\tanh^2(kh)) - (F+k\partial_k F) (\Omega^2\tanh(kh) + gk) \\ - kF\bigl(2\Omega\Gamma\tanh(kh) + \Omega^2(1-\tanh^2(kh))h + g\bigr) = 0,
\end{multline}
where $\Omega$ was defined above and $\Gamma := it\partial_tc_g + ic_g - 4\nu k$.

Integro-differential equations (\ref{eq:omega}) and (\ref{eq:group}) are too complex for any mathematical analysis. Therefore, we use numerical methods in order to explore some properties of the solutions.

\section{Numerical results and discussion}\label{sec:num}

In order to have an insight into the transient behaviour of the phase and group velocities, we have to solve numerically equations (\ref{eq:omega}) and (\ref{eq:group}) correspondingly. Hence, in this study we discretize all local terms with a first order implicit scheme, while the integral term $F(\omega,t)$ is computed in explicit way for the sake of computational efficiency. Resulting algebraic equations are solved analytically. Initial conditions were chosen according to \citep{Dias2007}:
\begin{equation*}
  \left.c_p\right|_{t=0} = \sqrt{\frac{g}{k}\tanh(kh)} - 2i\nu k, \quad
  \left.c_g\right|_{t=0} = \Bigl(\frac12 + \frac{kh}{\sinh(2kh)}\Bigr)
  \Re \left.c_p\right|_{t=0} - 4i\nu k.
\end{equation*}
The values of all parameters used in numerical computations can be found in Table \ref{tab:params}. The time step $\Delta t$ is chosen to achieve the convergence up to graphical resolution.

In this section upper and lower images always refer to the real and imaginary parts respectively unless special indications are given. Presented here results are normalized by $\sqrt{gh}$. Thus, all real parts at the infinitely long wave limit $kh\to 0$ take the unitary value.

\begin{table}
	\begin{center}
		\begin{tabular}{ccc}
  		\hline\hline
  		\textit{parameter} & \textit{definition} & \textit{value} \\
      \hline
        $\nu$ & eddy viscosity & $10^{-3}$ $\frac{m^2}{s}$ \\
      \hline
        $g$ & gravity acceleration & $9.8$ $\frac{m}{s^2}$ \\
      \hline
       $h$ & water depth & $1$ $m$ \\
      \hline
       $\Delta t$ & time step & $0.05$ s \\
      \hline\hline
		\end{tabular}
		\caption{Values of the parameters used in the phase and group velocities numerical computations.}
		\label{tab:params}
	\end{center}
\end{table}

\begin{figure}
	\centering
		\includegraphics[width=0.99\textwidth]{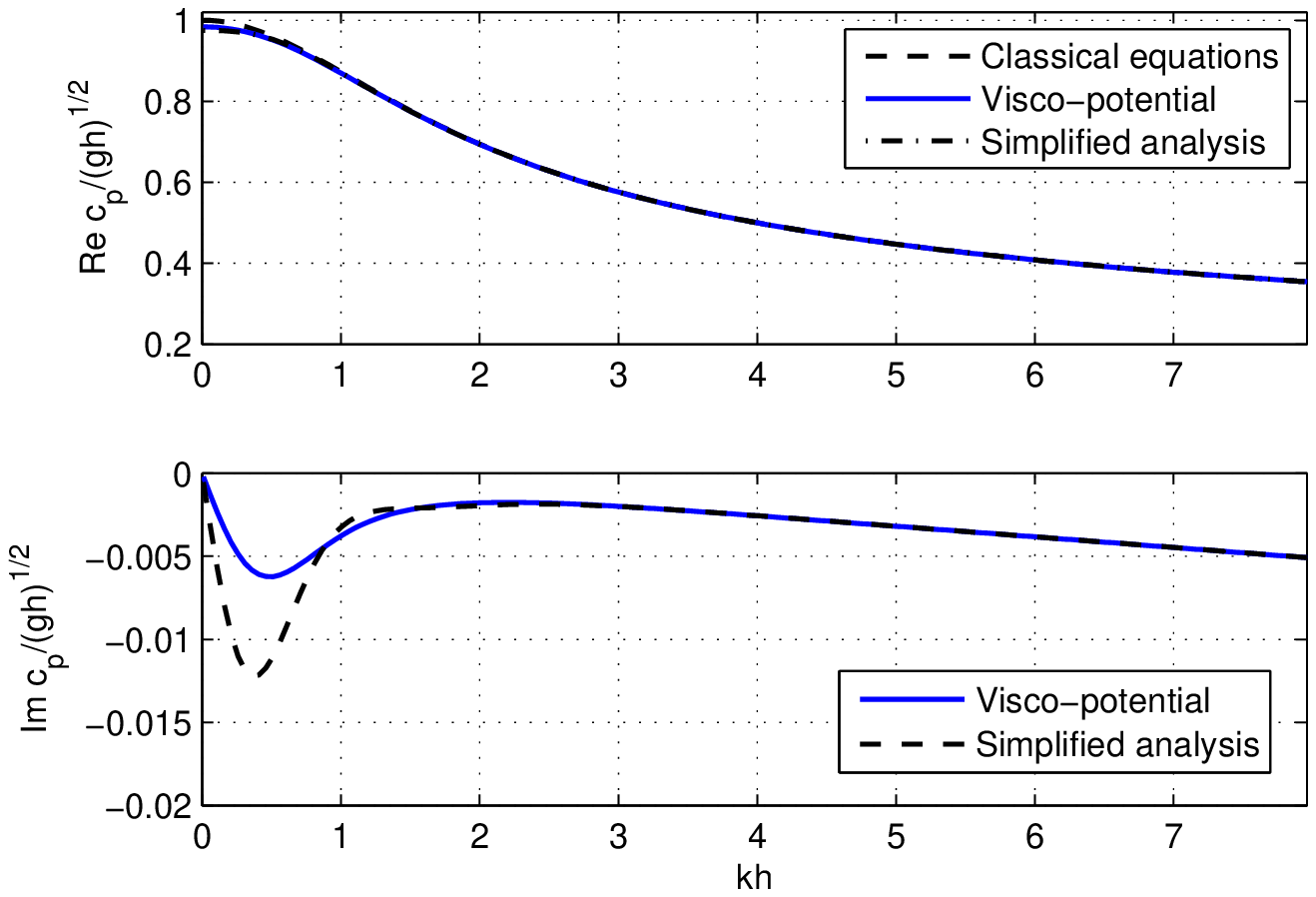}
	\caption{Real and imaginary parts of the phase velocity at $t=2$ s.}
	\label{fig:phase_t_4}
\end{figure}

\begin{figure}
	\centering
		\includegraphics[width=0.99\textwidth]{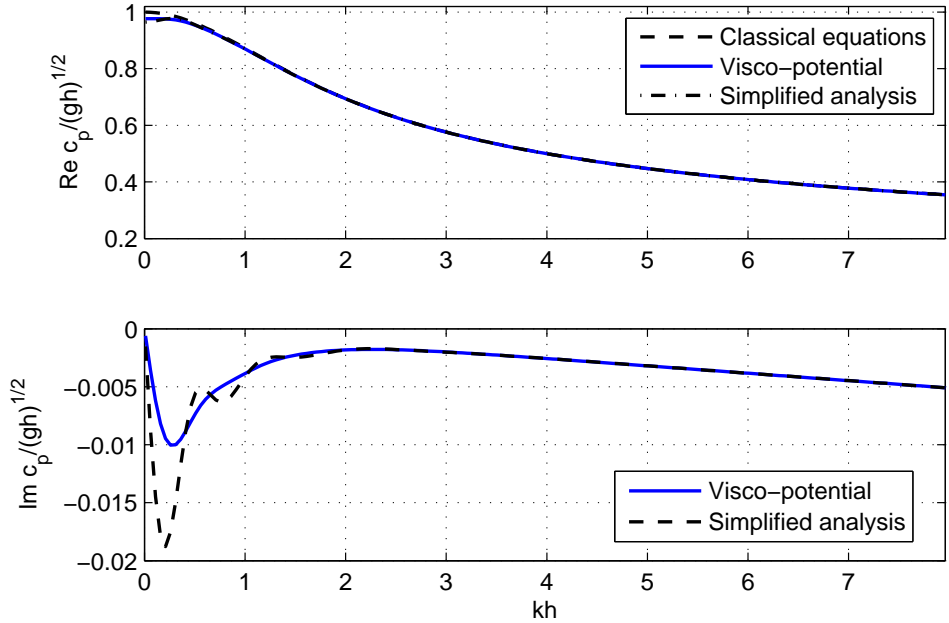}
	\caption{Real and imaginary parts of the phase velocity at $t=4$ s.}
	\label{fig:phase_t_8}
\end{figure}

First of all, on Figures \ref{fig:phase_t_4} and \ref{fig:phase_t_8} we show two snapshots of the phase velocity at different times. On these Figures we plot also computational results of our previous \textit{weakly dynamic} analysis \citep{Dutykh2008a}, which consists in neglecting differential evolution terms $it\pd{\omega}{t}$ in equation (\ref{eq:omega}). We can constate an excellent agreement for the real part of the phase velocity. However, with the simplified approach we get only the qualitative behaviour of the imaginary part. This discrepancy is located in the region of long waves where the boundary layer damping is predominant. The dissipation of short waves is perfectly described again. From quantitative point of view, our previous analysis overestimates the dissipation rate for long wavelengthes. We refer to \citep{Dutykh2008a} for the discussion and physical interpretation of the phase velocity behaviour.

\begin{figure}
	\centering
		\includegraphics[width=0.99\textwidth]{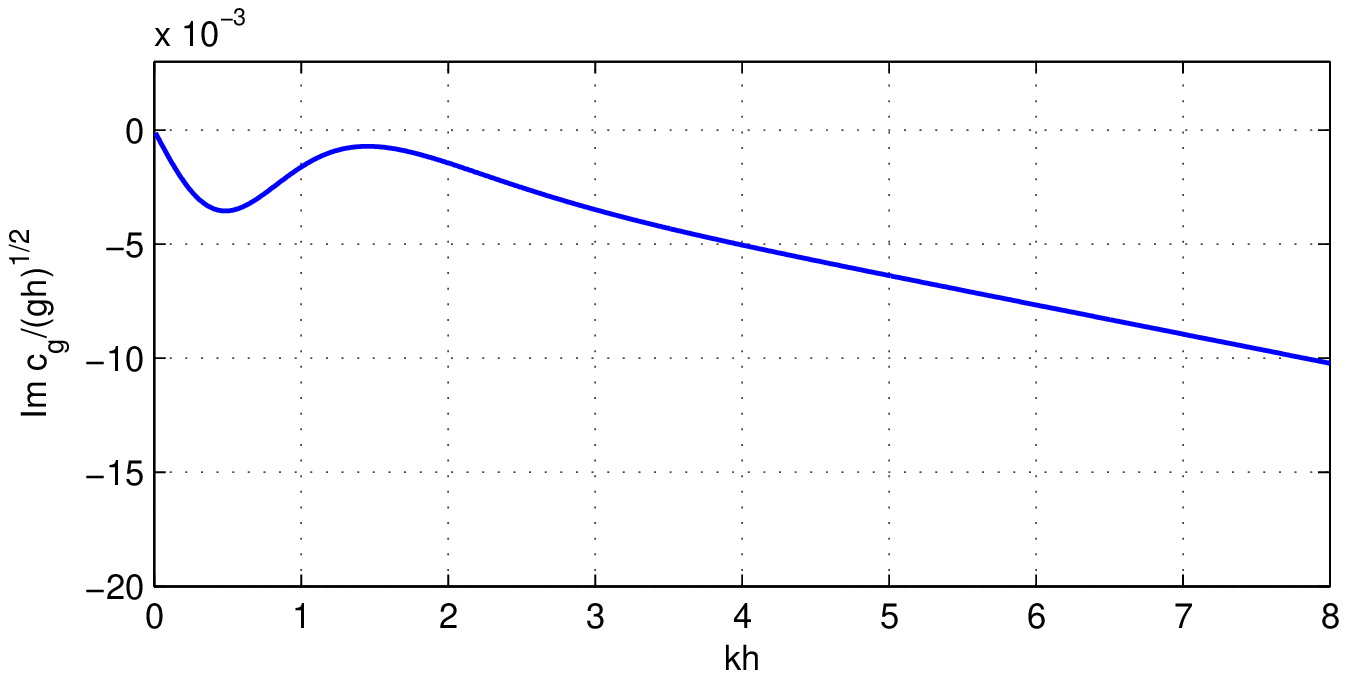}
	\caption{Imaginary part of the group velocity at $t = 1$ s.}
	\label{fig:group_t_08}
\end{figure}

\begin{figure}
	\centering
		\includegraphics[width=0.99\textwidth]{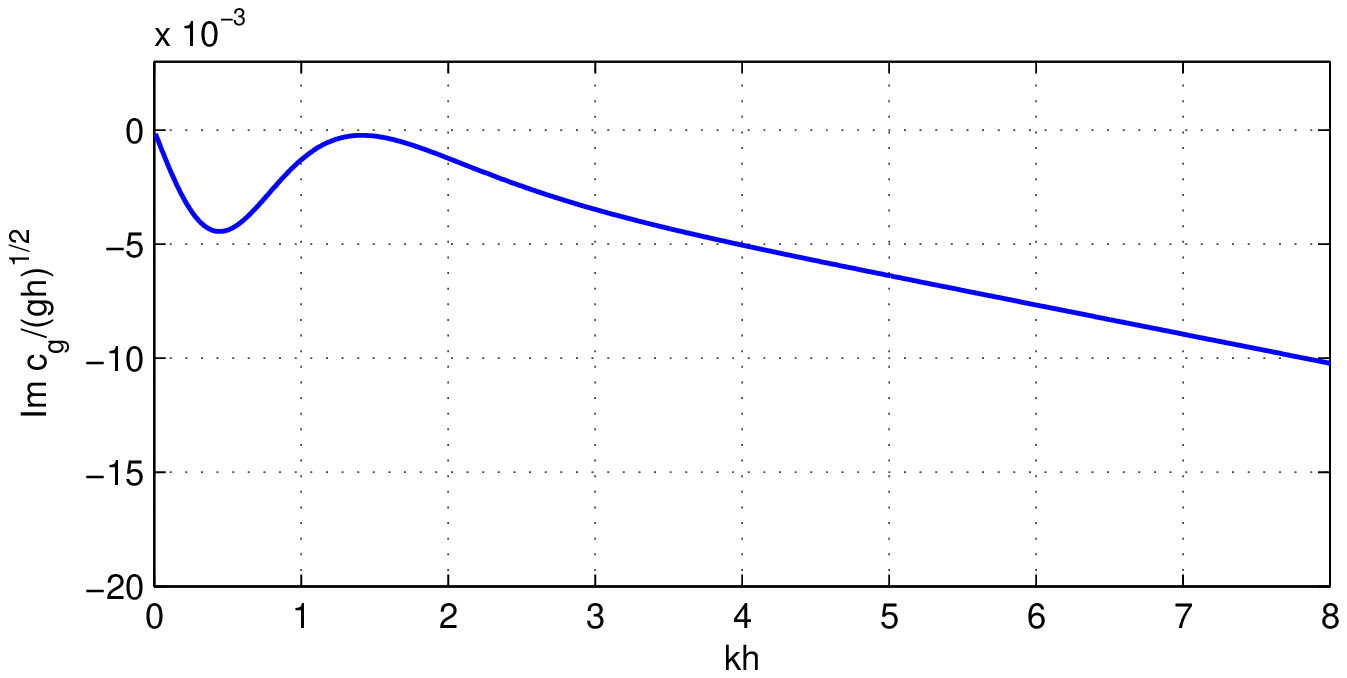}
	\caption{Imaginary part of the group velocity at $t = 1$ s.}
	\label{fig:group_t_1}
\end{figure}

\begin{figure}
	\centering
		\includegraphics[width=0.99\textwidth]{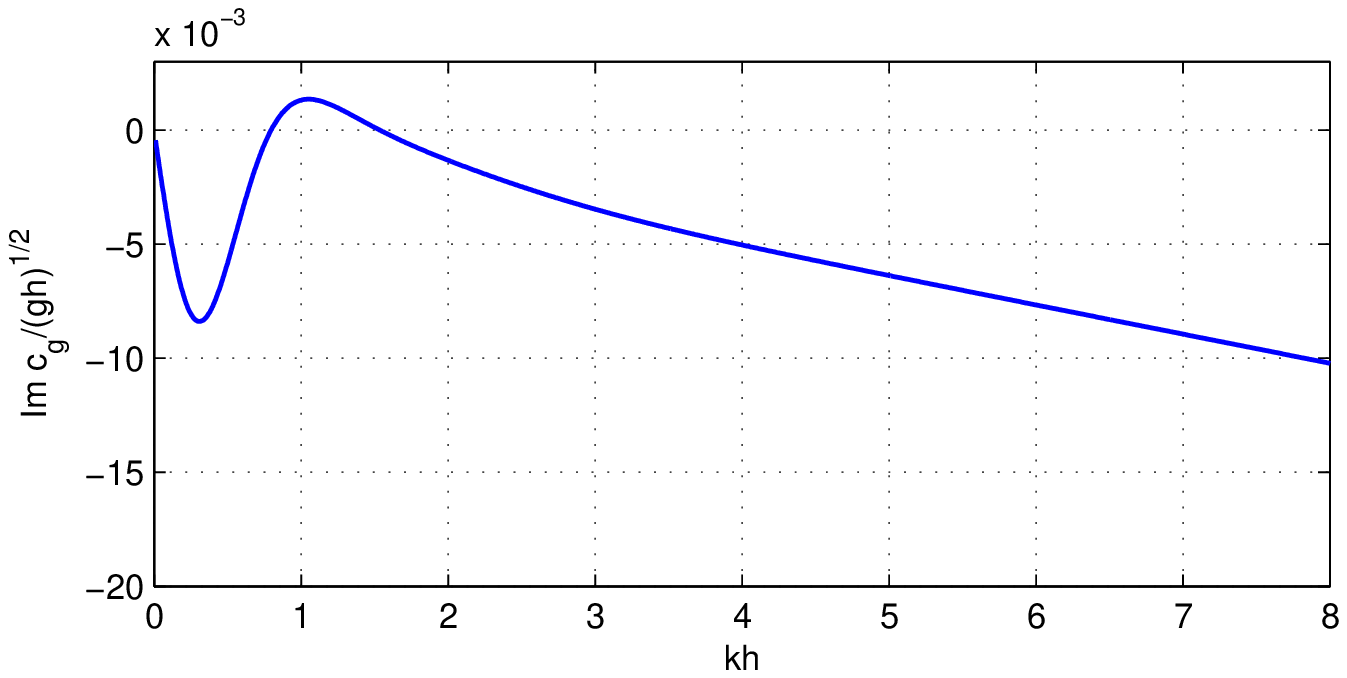}
	\caption{Imaginary part of the group velocity at $t = 2$ s.}
	\label{fig:group_t_2}
\end{figure}

\begin{figure}
	\centering
		\includegraphics[width=0.9\textwidth]{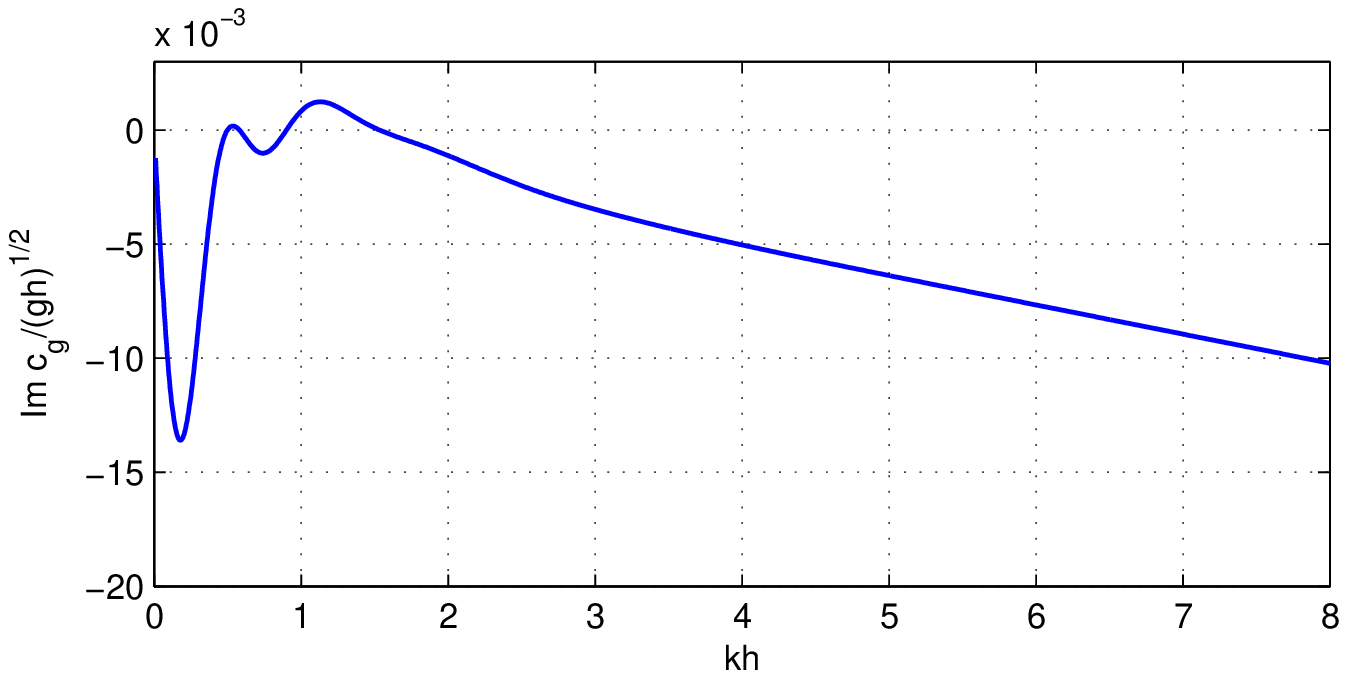}
	\caption{Imaginary part of the group velocity at $t = 4$ s.}
	\label{fig:group_t_4}
\end{figure}

\begin{figure}
	\centering
		\includegraphics[width=0.99\textwidth]{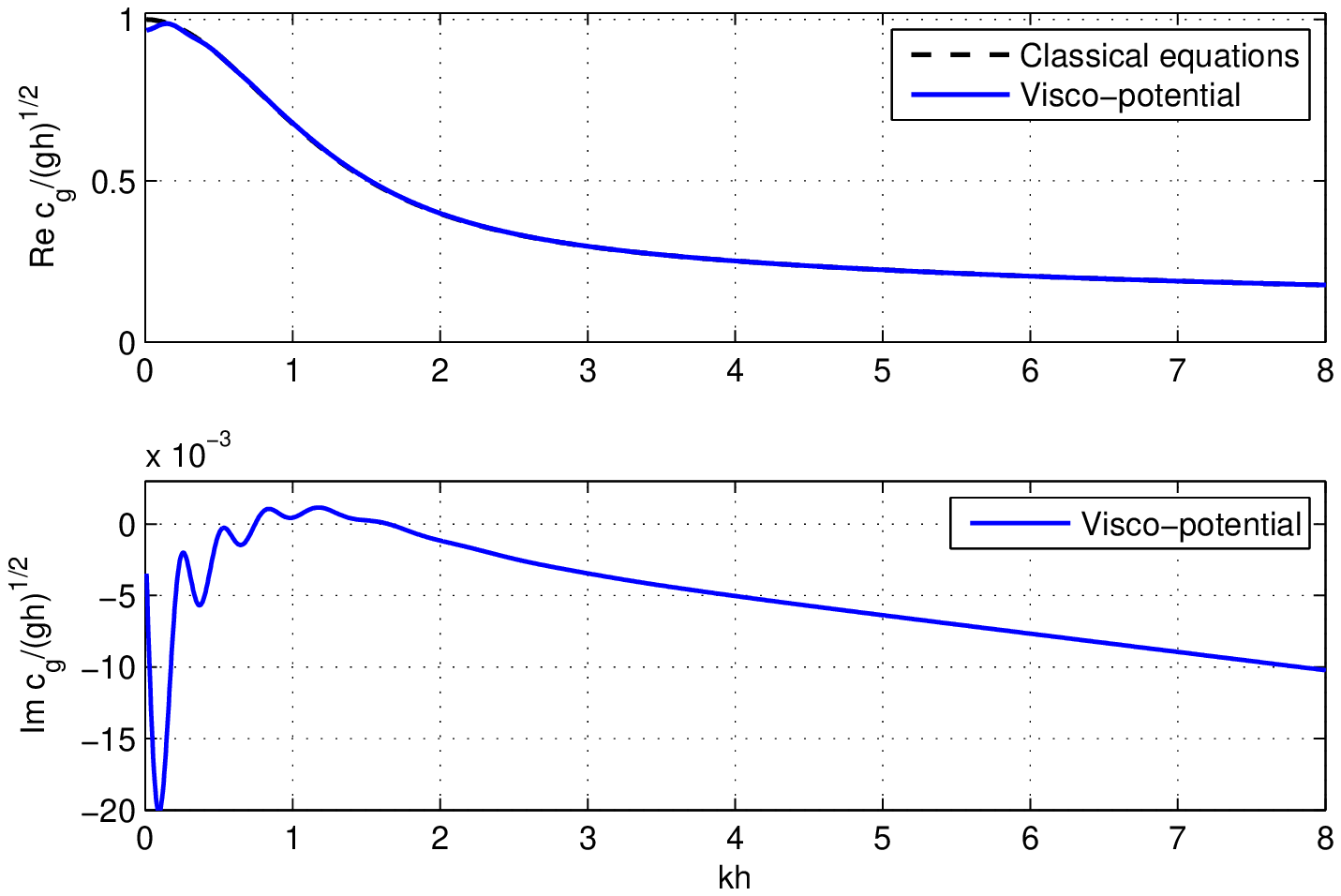}
	\caption{Real and imaginary parts of the group velocity at $t=8$ s.}
	\label{fig:group_t_8}
\end{figure}

Group velocity snapshots are presented on Figures \ref{fig:group_t_08} -- \ref{fig:group_t_8}. The real part of the group velocity $c_g$ is shown only on Figure \ref{fig:group_t_8} since its evolution is slow as in the phase velocity case. The whole animation of the group velocity can be downloaded at \citep{GRPanimation}.

The imaginary part of the group velocity remains negative until about $1$ s (see Figure \ref{fig:group_t_1}). Then, it was very unexpected for the author to observe the formation of modes with positive imaginary part (see Figure \ref{fig:group_t_2}). These destabilizing modes may be responsible of wavepackets disintegration. We refer to the next section for the discussion of this interesting result.

\section{Conclusions}\label{sec:concl}

In this work we performed an analysis of linear dispersion relation for visco-potential flow formulation. Several numerical snapshots of the phase and group velocities were presented. The main particularity in this study is that we make no assumption of the weak dependence on time. Therefore, we come up with additional terms of the form $it\omega'(t;k)$ which complicate analysis and numerical computations. We can conclude that we obtain a good qualitative agreement with preceding results \citep{Dutykh2008a}. Thus, we validate in some sense approximations made anteriorly. However, our computations reveal less oscillatory behaviour in the imaginary part of the phase velocity for intermediate wavelengths.

Furthemore, we studied the evolution of the group velocity with time for the visco-potential formulation. Physically, this quantity represents the energy propagation speed. There exists also another interpretation. The group velocity $c_g(t;k)$ can be seen as the wavetrain amplitude variation speed. Numerical simulations show the appearance of modes with positive imaginary part for $\frac12 < kh < \frac32$ in the group velocity. These modes may destabilize wave packets. Physically, this effect comes from the bottom boundary layer. Previous numerical study \citep{Wu2006} did not reveal this instability since the authors did not take into account for the boundary layer effect. Hence, they included only a local dissipative term which has a stabilizing effect.

It is well known that Stokes wavepacket can be disintegrated by the Benjamin-Feir instability \citep{Benjamin1967, Benjamin1967a}. This instability comes from the resonant quartet wave interaction. The classical result of T. Brooke Benjamin \citep{Benjamin1967}, refined later by R.S. Johnson \citep{Johnson1977}, states that no instability is present if $kh < \sigma_0$ and the  modes inside a certain band-width grow without bound if $kh \geq \sigma_0$. The critical value of $\sigma_0$ is estimated about $1.363$. There is a certain surprising interplay with our results that we would like to point out here. Our theory gives less sharper results. That is why we prefer to speak about the instability zone defined as $\U(t) = \set{kh\in \R^+: \Im c_g(t;k) > 0}$ in $kh$-space. According to our computations, the topological structure of the set $\U(t)$ is a union of finite number of intervals (see the animation \citep{GRPanimation}). It is curious that the critical value $\sigma_0$ lies inside this zone (see Figure \ref{fig:group_t_2}).

The wave turbulence can be perfectly described in the framework of potential flows. It gives direct cascade of energy from the waves spectral maximum to the dissipative region \citep{Zakharov1992}. However, in the present study we did not investigate the effect of viscous interaction with bottom onto the wave turbulence. Presumably, it will modify the dissipation region. This question deserves a thorough consideration.

Recently it was shown that the Benjamin-Feir instability can be enhanced by dissipation \citep{Bridges2007}. The peculiarity lies in the form of dissipative terms that we put into the model. Our computations show the stabilizing effect of local dissipative terms \citep{Segur2005, Wu2001, Wu2006} and the destabilizing effect of our nonlocal dissipation.

The experiments should confirm our theoretical result, provided that a generated wavetrain have sufficiently big wavelength to interact with the bottom boundary layer.

\section*{Acknowledgment}

The author would like to thank the Reviewer \#2 of the previous paper \cite{Dutykh2008a} for giving interesting ideas which lead to the results presented in this publication.

Moreover, the author would like to express his gratitude to Professors Fr\'ed\'eric Dias and Jean-Claude Saut for helpful and motivating discussions on visco-potential flows.

\bibliographystyle{alpha}
\bibliography{biblio}
\end{document}